# Development and Application of Nanocomposite Materials in Telecommunication Networks

Oleg Yaremko
*Lviv Polytechnic National University*
Lviv, Ukraine
oleh.m.yaremko@lpnu.ua

Nazariy Andrushchak
*Lviv Polytechnic National University*
Lviv, Ukraine
nandrush@gmail.com

Volodymyr Adamiv
*Institute of physical optics*
Lviv, Ukraine
vol.adamiv@gmail.com

Taras Rosa
*Lviv Polytechnic National University*
Lviv, Ukraine
taras1rosa@gmail.com

Monika Lelonek
SmartMembranes GmbH
petra.goering@smartmembranes.de

Anatoliy Andrushchak
*Lviv Polytechnic National University*
Lviv, Ukraine
anatolii.s.andrushchak@lpnu.ua

*Abstract* — **This article considers the issue of the development, production, and application of nanocomposite materials in telecommunication networks for the optical and subterahertz wave ranges. The results are obtained on the base of growing the nanocomposite materials from $Al_2O_3$ membranes and KDP crystals in a saturated solution.**

## I. Introduction

The development of modern technologies opens up new opportunities for bringing together achievements in various fields of production and research. The transmission of multimedia data has led to the emergence of networks with standards of 100G and 40G Ethernet. Modern wireless telecommunication networks have practically completely used both licensed and unlicensed bandwidths. To provide high-speed data transmission over the radio channel, the use of multi-position modulation was not enough. Developers of receiving and transmitting equipment increasingly focus on the sub-terahertz and terahertz bands (from 60 GHz to 3000 GHz), which does not require licensing. Until recently, there was no technological basis for the release of such devices. The specificity of the radio waves propagation in this band entails several limitations related to the speed of the stream and the weather conditions. Significant attenuation of the radio waves in the air reduces the receiving range. The requirements for antennas are also increasing, due to the main lobe position stabilization. The advantages include reduction of the interference and diffraction level, because of decreasing the receiving range possible increase in the number of stations in the cell, reducing the size of antennas. The presented problems point to the need to develop new materials for antennas, modulators, and converters using nanotechnology.

## II. Transport Distribution Networks With Application in the SubTerahertz Band

One of the promising application areas of sub-terahertz and terahertz technologies is telecommunication systems [1, 2]. In particular, the leading scientific and technical schools in the world are creating the sub-terahertz and terahertz band devices with fundamentally new dimensions, noise immunity and energy efficiency for 4G/LTE and 5G mobile communication networks, high-speed transmission of high and ultra-high definition video signals, radio relay systems for direct visibility and radars and sensors to obtain more accurate and detailed operational information about the state of the controlled object or terrain.

The use of technology radio over fiber (RoF – Radio over Fiber) opens up new possibilities [2]. Technologically, it is expedient to transmit radio signals over long distances through fiber-optic lines. Typically, a millimeter-wave radio signal is imposed on an optical carrier using an optoelectrical modulator and is put through an optical cable to a station where it is demodulated and radiated by an antenna. This technology allows extending the frequency range and improving electromagnetic compatibility, and using QPSK (Quadrature Phase Shift Keying), QAM (Quadrature Amplitude Modulation) and OFDM (Orthogonal frequency-division multiplexing) allow increasing spectral efficiency.

As modulators, the Mach-Zehnder modulator is used, and photonic crystals, made based on lithium niobate ($LiNbO_3$), having high speed, as well as films of barium titanate. Significant changes occur in the antenna feed chain. The use of composite materials reduces their size and allows the development of broadband antennas with a high gain.

The nanoporous matrices formation depends on the technological production process, which provides periodicity of the structure, open or close nanopores bottom and their form. The matrices material determines not only the range of application, optical or radio but also linear or non-linear properties, the suitability of this or that crystal filling method.

To conduct research on optical properties, several requirements regarding the geometric dimensions and surface processing cleanliness must be met. As a result, the formed structures can be used as an active medium in the optical range [3, 4]. Thus, the nanocomposites deserve attention in the quasi-optical range, whereby introducing heterogeneities, for example, changes in dielectric conductivity, the formation of slowing structures is possible, which creates possibilities for the antennas and filters development. Accordingly, the water-soluble crystalline ferroelectric KDP and the nanoporous matrices grew from aluminum by the anodizing method are used in the proposed research.

## III. Nanoporous Matrices Creation Technology Based on $Al_2O_3$

The active study of anodized aluminum oxide (AAO) began in the middle of the last century and remained an actual issue. At first, AAO was used in industry, as the basis of protective coatings. In addition to its mechanical characteristics, this material is primarily attractive because depending on the conditions it is possible to obtain orderly porous structures with given properties. That is, by changing the anodizing modes and variants of further structures modification, it is possible to obtain AAO with different characteristics and properties [5].

There are various methods for obtaining ordered AAO structures. All of them, one way or another, are reduced to anodizing the surface of the aluminum. To get oxide layers under controlled conditions, anodic oxidation (anodization) is used. An electrochemical process provides the formation of oxide on the surface of metals (including aluminum) and semiconductors due to anode polarization in an oxygen-containing aqueous environment with ion conductivity [6].

The resulting solution of weakly soluble acids AAO significantly differs in thickness, which can be achieved by parameters that allow controlling the growth of the thickness. It is known that except the temperature of the electrolyte, the growth of barrier oxide is limited only by the value of the applied voltage [7]. At the same time, the thickness of the porous layer depends on the current density and time.

The permissible maximum voltage value for the barrier layer is limited to the magnitude of the aluminum oxide breakdown voltage. The maximum thickness for such films is achieved at 500–700 V, which corresponds to 700–1000 nm. The thickness of the porous film may reach a much larger size than the first one. Here the temperature is also an important parameter. With the temperature increasing, the current density is also increasing at a constant voltage. At low temperatures (0–5 C) an oxide coating is formed, moreover, its support allows to achieve higher ordering of structures [8], oxide hardness [9] and change the pore shape [10-12].

For membrane manufacturing, the corresponding workpieces are cut from a high-purity aluminum sheet (99,999). The workpieces are cut by appropriate forms, which are subsequently used for various chemical manipulations to obtain a nanoporous structure. The workpieces must be thoroughly cleaned, as it may prevent further electrochemical processing of membrane manufacturing. Cleaning of aluminum workpieces is carried out by washing in acetone, isopropyl alcohol, and deionized water.

To obtain good surface cleanliness and eliminate various defects, electrochemical polishing is carried out. For aluminum polishing, a solution of $HClO_4$ and $C_2H_5OH$ are used in a ratio of 1:4. The polishing is carried out in the same forms as the subsequent processing to obtain the membranes (see Fig. 1, made in the premises of the SmartMembranes GmbH). The polishing process takes place at room temperature for 15-20 min. by using platinum electrodes and applying a voltage of 20 V.

After the first step of anodization, the oxide layer is removed, leaving patterned aluminum. Oxide removal is performed in a solution of $CrO_3$ and $H_3PO_4$ at 30 °C for approx. 10–12 h. The duration of the second step is dependent on the intended oxide layer thickness. Oxide growth rate depends on the solution and current, but typical values at steady state current are 2,5 µm/h for oxalic, 5 µm/h for sulfuric and 6 µm/h for phosphoric acid.

After the approximated anodizing time is achieved, the oxide layer thickness is measured using coating thickness meter. If the thickness is not sufficient, anodization is continued. When anodization is completed, aluminum layer under the oxide is removed using a water solution of $CuCl_2$ and HCl. As the reaction with aluminum is highly exothermic, a solution must be cooled using ice.

Oxide layer after aluminum removal has pore matrix on one side and penetrating through most of its thickness but is closed on the side contacting aluminum. To produce a membrane

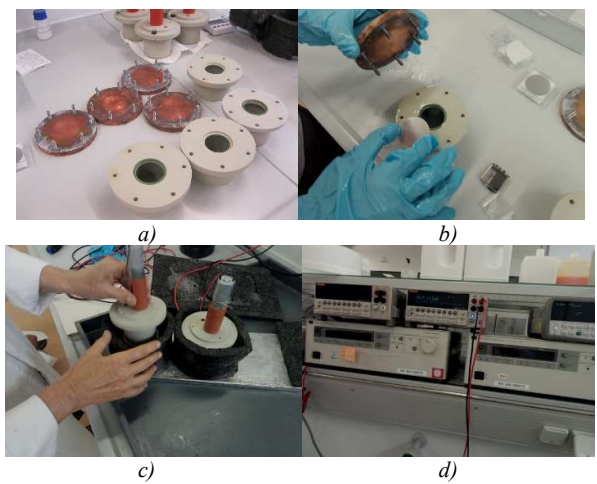

*a)* *b)*

*c)* *d)*

Fig. 1. Equipment used in the membranes $Al_2O_3$ manufacture: a) glasses with gaskets and copper supports for connection; b) placement of an aluminum plate and its fastening; c) connection of mixers and contacts to power supply units; d) power supplies and voltage and current control units.

pore bottoms must be opened. To do this, all membranes are floated on the surface of 10 % $H_3PO_4$ at 30 °C (Fig. 2). As soon as pore bottoms are open, the liquid can be seen on the upper surface of the membrane, and it can be removed. After that, the membranes thoroughly rinse with deionized water and dry.

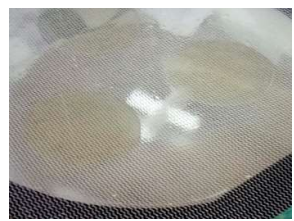

Fig. 2. An example of pores opening in AAO matrix

When the quality of the membranes is confirmed they are cut to desired final dimensions with a laser cutter.

IV. FILLING THE $AL_2O_3$ MEMBRANES WITH KDP CRYSTALS

A water bath with a temperature regulator was set up on a magnetic stirrer for the crystals growth from the aqueous solution at our Center of excellence for Innovative Technologies and Nanoengineering (Fig. 3). Magnetic stirrer and temperature regulator allowed to quickly and efficiently obtain a saturated solution of the corresponding salt (KDP), and to regulate the temperature regimes of nanocrystals growth in porous $Al_2O_3$ membranes.

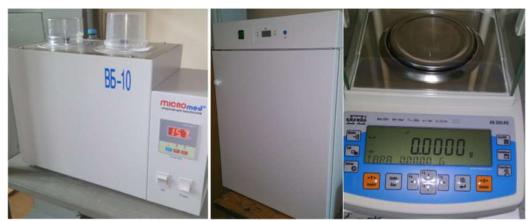

Fig. 3. The water bath installed on a thermal plate with a magnetic stirrer

The KDP (KH$_2$PO$_4$) crystals growth in the pores of the Al$_2$O$_3$ membrane was carried out on two types of membranes: with a pore diameter of 40 nm and 75 nm with an interpore distance of 125 nm. The membranes thickness was 100 μm.

The membranes were installed for crystals growth in filter cups (ceramic filter implanted in cylindrical glass) for filtration of the saturated solution (depriving of additional growth centers) from which nanocrystals grow (Fig. 4).

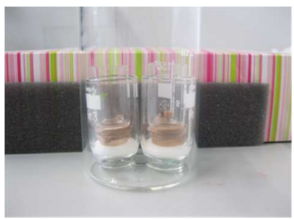

Fig. 4. The ceramic filter (white color below) implanted into a cylindrical glass

Following the recommendations from [13], the process of KDP crystals growth in nanoporous Al$_2$O$_3$ membranes is:

1. The water bath for KDP crystals growth is installed (Fig. 3). The crystals were grinded and placed in the glass to dissolve. A water bath is placed on a magnetic stirrer with controlled heating. The temperature is set to 57 °C. The temperature variation is 57-61°C.

2. The solution without crystals is obtained. The temperature is decreased to 55 °C to get a saturated KDP solution. The temperature variation is 55-57 °C.

3. The temperature is set to 54 °C. The temperature variation is 54-55 °C. A medical needle with a 1 mm hole was installed in the glass with the KDP solution for uniformly slow evaporation – increasing the solution saturation and the crystals growth.

4. The temperature variation is 54-55 °C. Solution evaporation in the glass is less than 1 mm/day. The first crystals formed – a sign of the solution saturation at the given temperature. The water bath is additionally sealed with duct tape. The temperature is not changed.

5. Two membranes (heated to the solution temperature) with the pore diameter of 40 nm were installed in separate containers (Fig. 4). All together closed and heated for 3 hours. The solution temperature decreased by 4 °C: from 54 °C to 50 °C. The formation of KDP crystals beyond the membranes is observed, which has significantly increased in size compared to the previous state.

6. Removing the first membrane. Rinsing it twice with deionized water, wiping dry and packing ($D$ = 40 nm, $L$ = 125 nm). The membrane exposition in the saturated KDP solution was 3 hours.

7. Removing the second membrane. Rinsing it twice with deionized water, wiping dry and packing ($D$ = 40 nm, $L$ = 125 nm). The membrane exposition in the saturated KDP solution was 24 hours.

8. After the growth manipulations and packing in separate containers, the membranes are stored in a desiccator with dried silica gel.

## V. Optical Measurements

After the manipulations above, a series of optical measurement was conducted to monitor the pores filling by KDP crystals on an FEI Versa 3D electron microscope (Fig. 5) with a resolution of 0.8 nm.

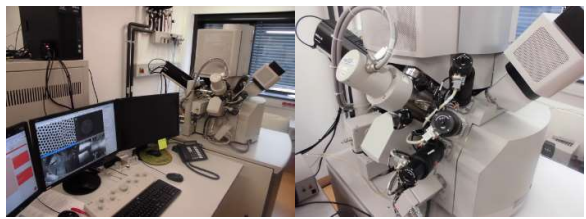

Fig. 5. FEI Versa 3D electron microscope

For complete control, two-sided and side view (at the membrane fracture) measurements were performed for both the clean membrane without crystals and the membranes with KDP crystals with a growth time of 3 hours and 24 hours. As can be seen in Fig. 6, the structure and shapes of the pores with a 40 nm diameter are visible. In Fig. 7, the partial filling of the membrane with KDP crystals is shown, where there are filled, filled partially and empty holes. Crystals are also can be seen on both surfaces of the membrane, which indicates the necessity to rinse and clean the surface with deionized water more thoroughly (with soaked cloth).

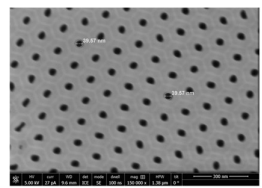

Fig. 6. The initial matrix: Al$_2$O$_3$ membrane with 40 nm pore diameter without the crystals

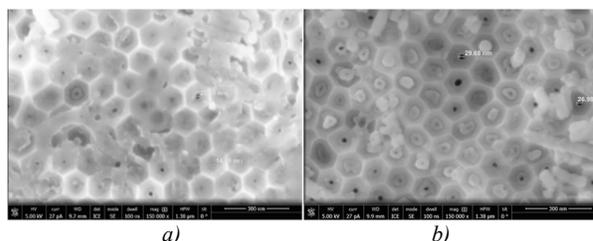

*a)*    *b)*

Fig. 7. The Al$_2$O$_3$ membrane with a 40 nm pore diameter and a 3 hours growth time in a saturated KDP solution *a)* the top side, *b)* the bottom side

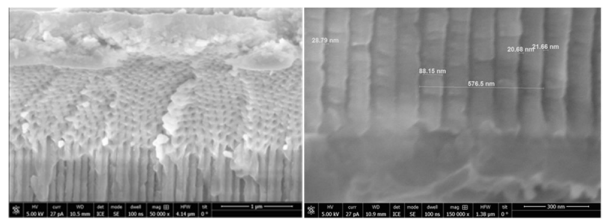

Fig. 8. The side view of the Al$_2$O$_3$ membrane with a 40 nm pore diameter and a 3 hours growth time in a saturated KDP solution

In Fig. 8 (side view – at the fracture) can be seen the incomplete filling of pore diameter and some redundancy of KDP on the membrane surfaces. Consequently, one should pay attention to the crystals growth technological process and its subsequent cleaning procedure.

The growth procedure – filling the Al$_2$O$_3$ matrix with a pore diameter of 40 nm for 24 hours in a saturated KDP

solution gave a slightly different result (Fig. 9-10). In these figures, it is shown a fundamental difference in the results obtained for membranes with 3 hours of crystal growth. Analyzing Fig. 7-10 it can be seen that the surface is almost completely covered by the crystalline KDP. However, the pores are partially filled. Consequently, an increase in the

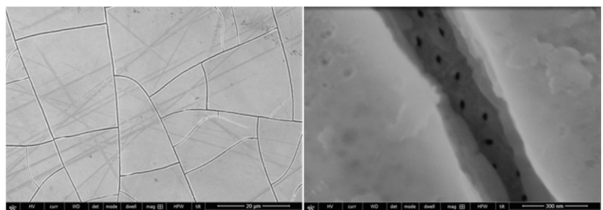
Fig. 9. The surface of the Al$_2$O$_3$ membrane with a 40 nm pore diameter and a 24 hours growth time in a saturated KDP solution

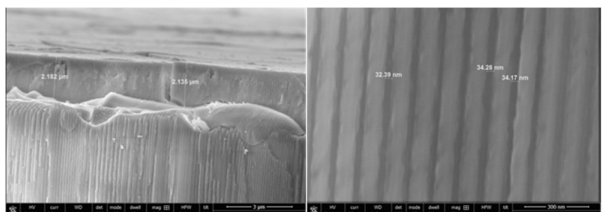
Fig. 10. The side view of the Al$_2$O$_3$ membrane with a 40 nm pore diameter and a 24 hours growth time in a saturated KDP solution

exposition time in the saturated solution negatively affects the filling process of the membranes – the KDP grows more on the surface than in the pores.

Similar growth procedures were performed with Al$_2$O$_3$ membranes with a pore diameter of 75 nm. Almost all of the described steps of the KDP crystals growth process in porous Al$_2$O$_3$ membranes are exactly repeated for membranes with a pore diameter of 75 nm. In Fig. 11-16 it is shown the growth results in membranes with a 75 nm pore diameter in a saturated KDP solution.

As can be seen from Fig. 11-16, the results are practically the same as for membranes with a 40 nm diameter. As can be seen from Fig. 11, the actual pore diameter of the initial matrix is not 75 but 70 nm. Fig. 12-13 also indicate the incomplete filling of pores with 3 hours growth time. Nanocrystals of different sizes are also visible on the surface.

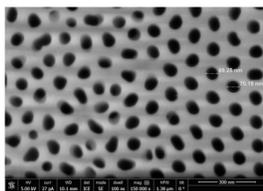
Fig. 11. The initial matrix: Al$_2$O$_3$ membrane with 75 nm pore diameter without the crystals. The remnant aluminum traces are visible. However, the diameter of the pores is not 75, but 70 nm

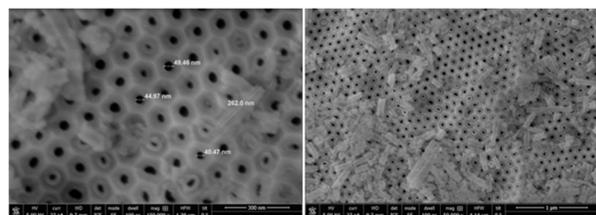
Fig. 12. The top side of the Al$_2$O$_3$ membrane with a 75 nm pore diameter and a 3 hours growth time in a saturated KDP solution

Figures 14-15 show that most of the surface of the Al$_2$O$_3$ membrane is covered by KDP, similar as in the case of 40 nm membranes. However, pores are partially filled with a crystalline KDP. Consequently, for the performed growth procedures in membranes with different pore sizes and at different growth duration, the results are practically the same. There is a great need for regulating pore filling by crystals. It is also necessary to minimize the closure of the membrane surface by crystalline or another component in the future.

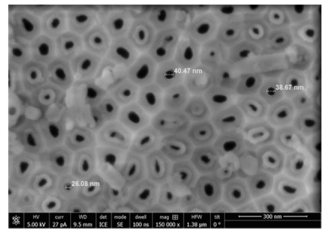
Fig. 13. The bottom side of the Al$_2$O$_3$ membrane with a 75 nm pore diameter and a 3 hours growth time in a saturated KDP solution

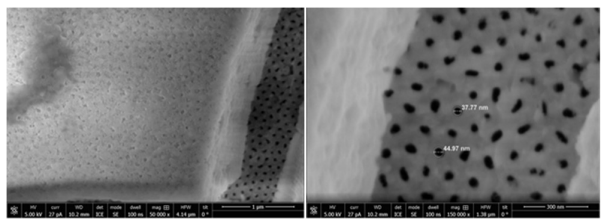
Fig. 14. The top side of the Al$_2$O$_3$ membrane with a 75 nm pore diameter and a 24 hours growth time in a saturated KDP solution

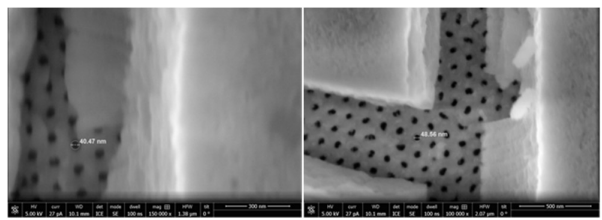
Fig. 15. The bottom side of the Al$_2$O$_3$ membrane with a 75 nm pore diameter and a 24 hours growth time in a saturated KDP solution

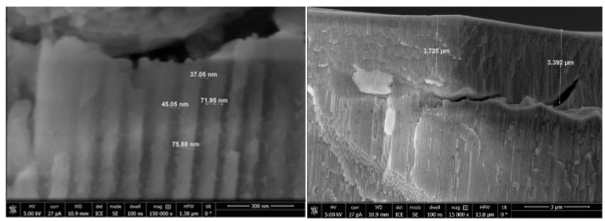
Fig. 16. The side view of the Al$_2$O$_3$ membrane with a 75 nm pore diameter and a 24 hours growth time in a saturated KDP solution

## VI. Transmittance Measurements

All the nanoporous matrices used in the proposed research was produced by the German company SmartMembrane Ltd. The diameter of pores were 40 and 75 nm, and the growth time for each particular case was 3 to 24 hours in the saturated KDP solution. The samples differed in the technology of growth and surface treatment of membranes. Samples 1 and 3 differ in duration and number of dives in the solution. Sample 2 was additionally polished after the immersion process. Sample 4, unlike the others, has a convex surface and surface damage due to mechanical deformations that arose in the manufacture was chosen because the heterogeneity of the surface allowed to consider the peculiarities of the formation of crystals.

To determine the presence of nanopores in the actual KDP crystals, the reflections measurements were performed in the infrared spectral range using a BRUKER IFS 66/S spectrophotometer. To do that the transmittance measurements, similar to the proposed in [14], were carried out both on the pure $Al_2O_3$ membrane (without KDP crystals), and on the membranes filled with KDP crystals. The measurement results are presented in Fig. 17-18.

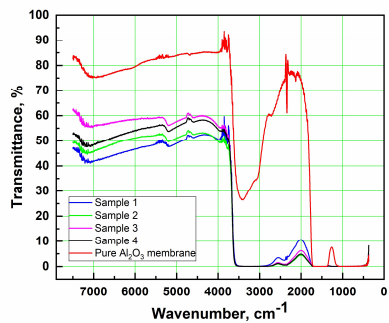

Fig. 17. Transmittance spectrums of different $Al_2O_3$ membrane samples

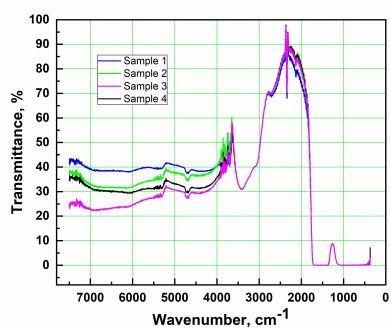

Fig. 18. Differential transmittance spectrums of $Al_2O_3$ membrane samples with KDP crystals

The spectrum of the pure and filled with KDP crystals membranes differs in the region of 1000-3500 $cm^{-1}$, which indicates the presence of other substances in the nanoporous.

The IFS 66/S spectrometer has a limit in the low-frequency range 55,000-1,500 $cm^{-1}$, which makes it impossible to see the maximum characteristic of the KDP crystals in the region from 1000 to 300 $cm^{-1}$. To prove the presence in the nanocomposites of KDP crystals, it will be necessary to conduct measurements on an X-ray diffractometer similar to the proposed in [15].

## VII. Conclusion

Using the premises of the Center of Excellence for Innovative Technologies and Nanoengineering the setup for growing the crystallites from a saturated KDP solution was created. Two growth cycles have been performed on filling nanoporous $Al_2O_3$ membranes with different pore diameters by crystalline KDP from a saturated solution.

Electron microscopy investigations were conducted to verify the quality of filled membranes of different diameters and growth time (see Fig. 6-16). To fill the nanopore with crystals, it is necessary to reduce the time of immersion in the solution and to conduct it repeatedly, to reduce the size of the crystallites on the surface.

Measurements of the transmittance have been performed in the infrared spectral range on both the pure $Al_2O_3$ membranes without KDP crystals and $Al_2O_3$ membranes filled with KDP crystals. In Fig. 17-18 it is shown in the region from 1500-3500 $cm^{-1}$ the presence of substances in nanoporous that are not presented in a pure membrane. This confirms the fact of filling the nanopores with the substance; however, to confirm the presence of KDP crystals, it is necessary to use devices with a dynamic range of 300-1000 $cm^{-1}$.

Acknowledgment

This result of the investigation is a part of a project that has received funding from the European Union's Horizon 2020 research and innovation programme under the Marie Skłodowska-Curie grant agreement No 778156. This work was also supported by the Ministry of Education and Science of Ukraine in the frame of the "Nanocrystallite" (#0119U002255) and "SubTera" projects (#0119U100609).